\documentclass[11pt,twoside]{article}
\usepackage{asp2006}
\usepackage{epsf}
\usepackage{graphics}
\usepackage{lscape}
\begin{document}
\title{Automatic Active-Region Identification and Azimuth
  Disambiguation of the SOLIS/VSM Full-Disk Vector Magnetograms}
\author{M. K. Georgoulis,\altaffilmark{1}
N.-E. Raouafi,\altaffilmark{2}
and C. J. Henney\altaffilmark{2}}
\altaffiltext{1}{Johns Hopkins University Applied Physics Laboratory 
(JHU/APL), Laurel, MD, USA}
\altaffiltext{2}{National Solar Observatory (NSO), Tucson, AZ, USA}
\begin{abstract}
The Vector Spectromagnetograph (VSM) of the NSO's Synoptic Optical
Long-Term Investigations of the Sun (SOLIS) facility is now
operational and obtains the first-ever vector magnetic field
measurements of the entire visible solar hemisphere. To fully exploit
the unprecedented SOLIS/VSM data, however, one must first address two 
critical problems: first, the study of solar active regions requires an
automatic, physically intuitive, technique for active-region
identification in the solar disk. Second, use of active-region 
vector magnetograms requires removal of the azimuthal
$180^o$-ambiguity in the orientation of the transverse magnetic field
component. Here we report on an effort to address both problems
simultaneously and efficiently. 
To identify solar active regions we apply an
algorithm designed to locate complex, flux-balanced, magnetic
structures with a dominant E-W orientation on the disk. Each of the
disk portions corresponding to active regions is thereafter extracted
and subjected to the Nonpotential Magnetic Field Calculation (NPFC)
method that provides a physically-intuitive solution of the 
$180^o$-ambiguity. Both algorithms have been integrated
into the VSM data pipeline and operate 
in real time, without human intervention. 
We conclude that this combined approach can
contribute meaningfully to our emerging capability for full-disk
vector magnetography as pioneered by SOLIS today and 
will be carried out by ground-based and space-borne magnetographs in
the future.  
\end{abstract}
\section{Introduction}
SOLIS is a state-of-the art ground-based facility dedicated to the
study of the Sun and its magnetic atmosphere for decades to come. 
It consists of three instruments - the vector spectromagnetograph
(VSM), the Integrated Sunlight Spectrometer (ISS), and the Full-Disk
Patrol (FDP). An overview on each instrument can be found in
Keller, Harvey, \& Giampapa (2003). The VSM, in particular, operates
from Kitt Peak since May 2004. Among other daily measurements, the VSM
performs complete Stokes polarimetry at the Fe {\small I} 
$630.2\;nm$ photospheric spectral line. 
An inversion of the Stokes images
provides full-disk vector magnetograms of the solar photosphere (see
Henney, Keller, \& Harvey, 2006, for details). The VSM data are
the first spectrographic full-disk vector magnetograms ever obtained. 
Partial-disk vector magnetography is performed by a
handful of ground-based instruments and, recently, by the
spectro-polarimeter of the Solar Optical Telescope (SOT) on board the
{\it Hinode} satellite (Lites, Elmore, \& Streander 2001). {\it Hinode}
carries the first space-based vector magnetographs, while the first
air-borne vector magnetograph was included in the Flare Genesis
Experiment balloon payload (Rust, 1994). The first 
space-based full-disk vector magnetograms will be acquired  
by the Helioseismic and Magnetic Imager (HMI; Scherrer 2002) 
on board the imminent Solar Dynamics Observatory (SDO). From the above
series of groundbreaking developments, one feels that solar vector
magnetography is probably entering its golden era. Therefore, it is
both essential and timely to address and efficiently solve the
core problems associated with it in order to fully exploit the
unprecedented existing and future vector magnetogram data. 

Vector magnetic field measurements inferred by the Zeeman effect
suffer from an intrinsic azimuthal ambiguity in the 
orientation of the transverse magnetic field component (Harvey,
1969). Indeed, the properties of the transverse Zeeman effect remain
invariant under the transformation $\phi \rightarrow \phi +\pi$, where
$\phi$ is the azimuth angle. The problem of the $180^o$-ambiguity has
proved notoriously difficult to solve self-consistently, despite 
numerous attempts (for an overview, see Metcalf et al., 2006, and 
references therein). For full-disk magnetograms, an additional problem 
appears in case active regions should be studied. How are
active regions to be extracted from the full-disk measurements? This
task may appear trivial if performed manually, but it is quite
daunting in case voluminous data are to be processed. 
Moreover, applications intimately related to
active regions and their magnetic evolution such as, for example, 
magnetic helicity calculations or major flare forecasting, require a
robust definition of active regions and their spatial extent. 

In the following, we discuss our proposed solutions to both active
region identification and azimuth disambiguation. Both techniques have
been integrated into the VSM data pipeline and are operating in real
time owing to the modest computational resources they require. 
Their operation is smooth and their solutions are robust, 
and this holds promise for automatic application to future
full- or partial-disk vector magnetogram data.
\section{VSM Data Processing}
\subsection{Automatic Identification of Solar Active Regions}
\begin{figure}[!ht]
\plotone{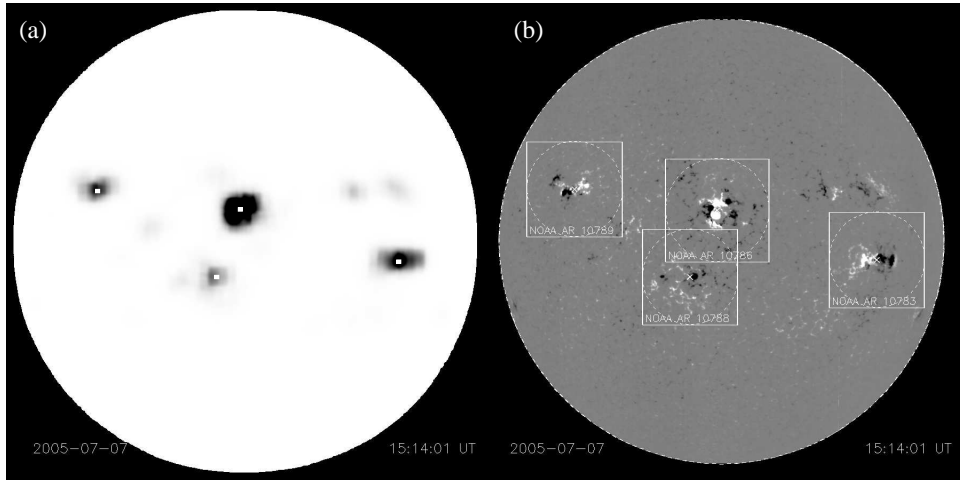}
\caption{\label{f1}
Automatic active-region identification in a full-disk VSM magnetogram
obtained on 2005 July 7, at around 15:14 UT. {\bf (a)} Negative of the
intensity image discussed in step 1.
The centroid location for each qualifying intensity enhancement 
(see steps 2 and 3) is indicated by a white dot. Notice that some minor
enhancements do not qualify for active-region association. 
{\bf (b)} The final result of
the identification process. Each active region is enclosed by a square
prescribed to an outer circle from each centroid, as described in
step (e). The corresponding NOAA numbers are also provided.} 
\end{figure}
Our active-region identification method 
does not require vector magnetograms. It 
was conceived by B. J. LaBonte and was first applied by
LaBonte, Georgoulis, \& Rust (2007) to full-disk line-of-sight
magnetograms of the Michelson-Doppler Imager (MDI - Scherrer et al.,
1995) on board the Solar and Heliospheric Observatory (SoHO). The
technique consists of five steps, namely: 
\begin{enumerate}
\item[1.] Smooth the full-disk line-of-sight magnetic field $B_{\ell}$
  using a smoothing window with linear size equal to one supergranular
  diameter (SGD), that is, $\sim 40\arcsec$, or $\sim 30\;Mm$ on the
  solar surface. 
  In the smoothed image, (i) test for bipolarity, enhancing
  magnetic polarity inversion lines, and (ii) calculate the gradient
  between the two polarities, emphasizing their E-W orientation. From
  this information (smoothed image, bipolarity, orientation) create an
  intensity image with enhancements corresponding to the active
  regions present on the disk (Figure \ref{f1}a). 
\item[2.] Coalign the intensity image with the actual
  magnetogram. Identify magnetic flux concentrations that coincide
  with the image's enhancements and discard those with flux imbalance
  larger than a prescribed threshold. 
\item[3.] For each qualifying intensity enhancement, determine an
  intensity-weighted centroid (barycenter). 
\item[4.] Match the location of each centroid with NOAA's Space
  Environment Center (SEC) archives to assign AR numbers to each
  centroid. For multiple numbers, choose the one with assigned
  location closest to the centroid. 
\item[5.] Starting from each centroid, determine the spatial extent
  (area) of the respective active region.
\end{enumerate}
Smoothing by 1 SGD provides a first clue of the
existing active regions and their locations on the disk. 
The size of the smoothing window reflects
the fact that supergranules are the fundamental convection cells
responsible for the large-scale magnetic fields in the Sun and,
therefore, for active-region formation 
(see, e.g., Leighton, Noyes, \& Simon, 1962). The flux imbalance 
criterion points to the fact that active regions are
mainly closed magnetic field structures. 
Nonetheless, the imbalance tolerance
limit is an external variable and can be set at will. The dominant E-W
orientation criterion helps avoid identifying accidental flux
associations as active regions. The centroid calculation helps 
associate the identified flux concentrations with the standard NOAA
active-region number database. 

The remaining task is the determination of the spatial extent of each
active region. This is accomplished in five steps, namely:
\begin{enumerate}
\item[(a)] Draw two concentric circles around each centroid - an inner
  one with radius equal to 1 SGD and an outer one
  with radius equal to 6 SGD. 
\item[(b)] Define circles with increasing radii from 1 to 6
  SGDs and find the average value of $|B_{\ell}|$ 
  on each perimeter.
\item[(c)] Find the median of the average $|B_{\ell}|$-values between
  5 and 6 SGDs from the centroid. Use this median
  as a $1-\sigma$ threshold. 
\item[(d)] Check the perimeter averages of $|B_{\ell}|$ and stop at
  the circle where the average $|B_{\ell}|$ falls below $2 \sigma$,
  i.e., becomes smaller than twice the above median.
\item[(e)] Add 1 SGD to the radius of this circle
  and draw the square prescribed to this radius. This square will 
  outline the edges of the identified active region (Figure \ref{f1}b). 
\end{enumerate} 
We underline that our size calculation technique, 
namely, defining an annulus with
internal and external radius of 5 and 6 SGDs,
respectively, to define the $1-\sigma$ threshold, precludes strong
magnetic fields from being included in areas belonging to multiple
active regions. Indeed, notice the overlapping between the areas of
NOAA active regions (ARs) 10786 and 10788 in Figure \ref{f1}b. This
common area {\it does not} include strong magnetic
fields. This stems from the fact that if the annulus cuts through
strong magnetic flux accumulations, then the  $1-\sigma$ threshold
will be higher and the average $|B_{\ell}|$ value will drop below 
$2 \sigma$ at shorter distances from the centroid, thereby imposing
a smaller spatial extent for the examined active region. We have
run numerous tests with full-disk magnetograms to verify that the
technique performs as expected in the vast majority of cases. 
%
%
\subsection{Azimuth disambiguation}
Our azimuth disambiguation method is the 
Nonpotential Magnetic Field Calculation (NPFC). 
The technique was introduced by Georgoulis (2005) and was
slightly refined (NPFC2) as discussed in Metcalf et al. (2006). In
this work, it was tested against several other techniques and was
found to reproduce successfully the correct azimuth solution in
highly nonpotential, noise-free, synthetic vector
magnetograms. Moreover, it was among the fastest azimuth
disambiguation techniques. In further comparisons, facilitated by a
series of Azimuth Resolution Workshops, the NPFC2 method 
successfully reproduced the correct solution 
in nonpotential synthetic vector magnetograms where various levels of
noise were embedded. Overall, the NPFC2 (hereafter NPFC) 
method demonstrated its speed and efficiency even in cases with
extreme noise levels, that eventually led to its integration to the 
VSM software package. 

\begin{figure}[!t]
\plotone{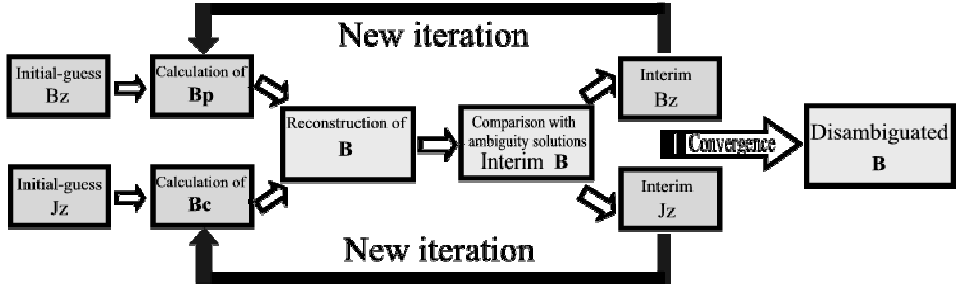}
\caption{\label{f2} A graphical description of the NPFC algorithm.}
\end{figure}
The physics behind the NPFC method is simple and starts by noticing
that any magnetic field configuration $\mathbf{B}$ can be decomposed
into a current-free (potential) component $\mathbf{B_p}$ and a
current-carrying component $\mathbf{B_c}$, i.e.
\begin{equation}
\mathbf{B} = \mathbf{B_p} + \mathbf{B_c}\;\;.
\label{eq1}
\end{equation}
If the magnetic structure is rooted in a lower boundary plane $S$
and extends in the half-space above it, then $\mathbf{B_p}$ and
$\mathbf{B_c}$ can be calculated on $S$ provided that 
the vertical (normal to $S$) component $B_z$ 
of the magnetic field $\mathbf{B}$ and the vertical
component $J_z$ of the electric current density $\mathbf{J}$
(where $\mathbf{J}$ is calculated by Ampere's law), 
respectively, are known on $S$. For $\mathbf{B_p}$
there are multiple calculation methods including Green's functions
(Schmidt, 1964) and Fourier transforms (Alissandrakis, 1981). For
$\mathbf{B_c}$, one utilizes the gauge conditions applying
to $S$, and especially the fact that $\mathbf{B_c}$ has only
an azimuthal component on $S$, i.e., $B_{c_z}|_S=0$. Assuming that 
$B_{c_z}$ does not vary significantly with height on $S$, i.e., 
$(\partial B_{c_z} / \partial z)|_S =0$, $\mathbf{B_c}$ can be
calculated in Fourier space and then inverted into Cartesian space,
i.e., 
\begin{equation}
\mathbf{B_c} = \mathcal{F}^{-1}[ {{i k_y} \over {k^2}} \mathcal{F} (j_z)] 
               \mathbf{\hat{x}} + 
               \mathcal{F}^{-1}[ {{-i k_x} \over {k^2}} \mathcal{F} (j_z)]
               \mathbf{\hat{y}}\;\;.
\label{eq2}
\end{equation}
In Equation (\ref{eq2}) we have $k^2 = k_x^2 + k_y^2$ for the harmonic
$(k_x, k_y)$ and $j_z = (4 \pi/c) J_z$, while $\mathcal{F}(r)$, 
$\mathcal{F}^{-1}(r)$ are the direct and inverse Fourier 
transforms of $r$, respectively, at $(k_x, k_y)$. Notice that the only
assumption of the NPFC method is  
$(\partial B_{c_z} / \partial z)|_S =0$. 
This assumption is quite reasonable because $B_{c_z}|_S=0$ and, for
small length elements, one might expect that  $B_{c_z} \simeq 0$
slightly above $S$, unless the magnetic field lines undergo dramatic
changes of orientation within the elementary height. 
Assuming that $S$ is the plane of the magnetic
field measurements, an unambiguous magnetic field
$\mathbf{B}$ can be reconstructed on this plane from $\mathbf{B_p}$
and $\mathbf{B_c}$. The problem, of course, 
is that both $B_z$ and $J_z$, required to calculate $\mathbf{B_p}$ and
$\mathbf{B_c}$, respectively, are 
subject to the $180^o$-ambiguity and are not known a priori. This
is where the numerical part of the NPFC method begins. 

Figure \ref{f2} provides a graphical representation of the numerical
process. Initial-guess distributions of $B_z$ and $J_z$ are used to
calculate the initial-guess $\mathbf{B_p}$ and $\mathbf{B_c}$,
respectively. The total field $\mathbf{B}$ is then reconstructed
through Equation (\ref{eq1}). The reconstructed field is compared to
the two equally possible ambiguity solutions and is set equal to the
solution that is closer to it, for each location of the measurements'
plane. This interim $\mathbf{B}$-configuration gives rise to new 
$B_z$- and $J_z$-solutions. From them, new $\mathbf{B_p}$- and 
 $\mathbf{B_c}$-configurations are produced, and the algorithm
proceeds to a new iteration. Convergence is judged by the number of
strong-field vectors whose orientation changes from one iteration to
the next. When no vectors (or a small number of vectors) are flipped
for 10 consecutive iterations, the process stops and the latest 
reconstructed $\mathbf{B}$-configuration is the suggested
disambiguation solution.

\begin{figure}[!t]
\plotone{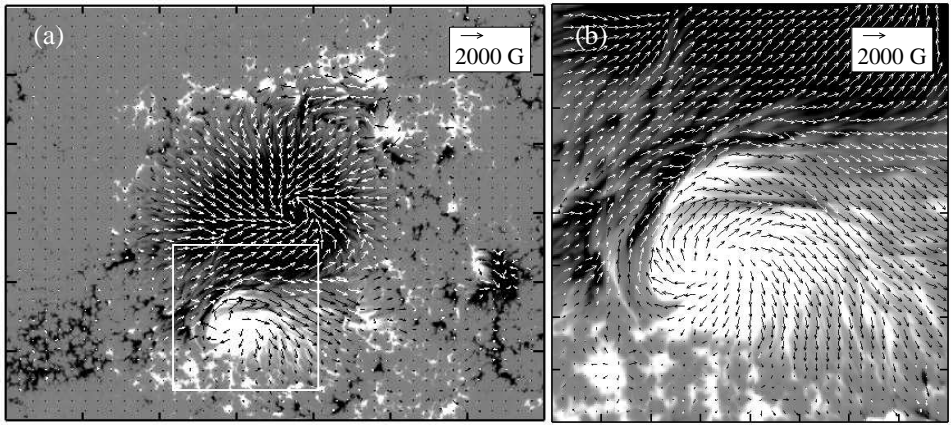}
\caption{\label{f3} Example of the NPFC disambiguation solution.
  Shown are the heliographic magnetic field components on the image
  plane. {\bf (a)} The solution on part of a {\it Hinode}/SOT
  spectro-polarimeter vector magnetogram depicting NOAA AR 10930, 
  obtained on 2006 December 11 between 13:53 UT and 15:15 UT.
  The data features a sunspot complex separated by a strongly sheared
  magnetic polarity inversion line, enclosed by the white square. 
  Tick mark separation is
  $20\arcsec$. {\bf (b)} The disambiguation solution for the enclosed
  area. Tick mark separation is $5\arcsec$. In both images, the
  vertical magnetic field saturates at $\pm 1000\;G$. Data courtesy of
  B. W. Lites.}     
\end{figure}
The entire calculation requires minimal computing resources and is
completed in a matter of minutes, even at an ordinary desktop
workstation. Remarkably, the speed of the convergence does {\it not}
depend on the complexity of the studied magnetogram but, rather, on
its quality and linear size. In other words, the quality of the solution
depends strongly 
on the quality of the measurements. To demonstrate this, we 
show in Figure \ref{f3} a disambiguated vector magnetogram from 
{\it Hinode's} SOT spectro-polarimeter that was 
inverted and kindly provided to us by B. W. Lites. 
These data are of exceptionally high seeing-free quality
and nearly unsurpassed spatial resolution ($\sim 0.16\arcsec$ per pixel). 
The depicted active region (NOAA AR 10930) shows a significant degree
of complexity, especially along a strongly sheared magnetic polarity
inversion line. The magnetogram's field of view had a very large
linear size, namely, $1904 \times 1024$ pixels. Yet the NPFC method
required only $\sim 20\;min$ and $34$ iterations to
converge. The respective computing time for typical linear 
dimensions of, say, $512 \times 512$ pixels, would not exceed $2\;min$. 

\begin{figure}[!t]
\plotone{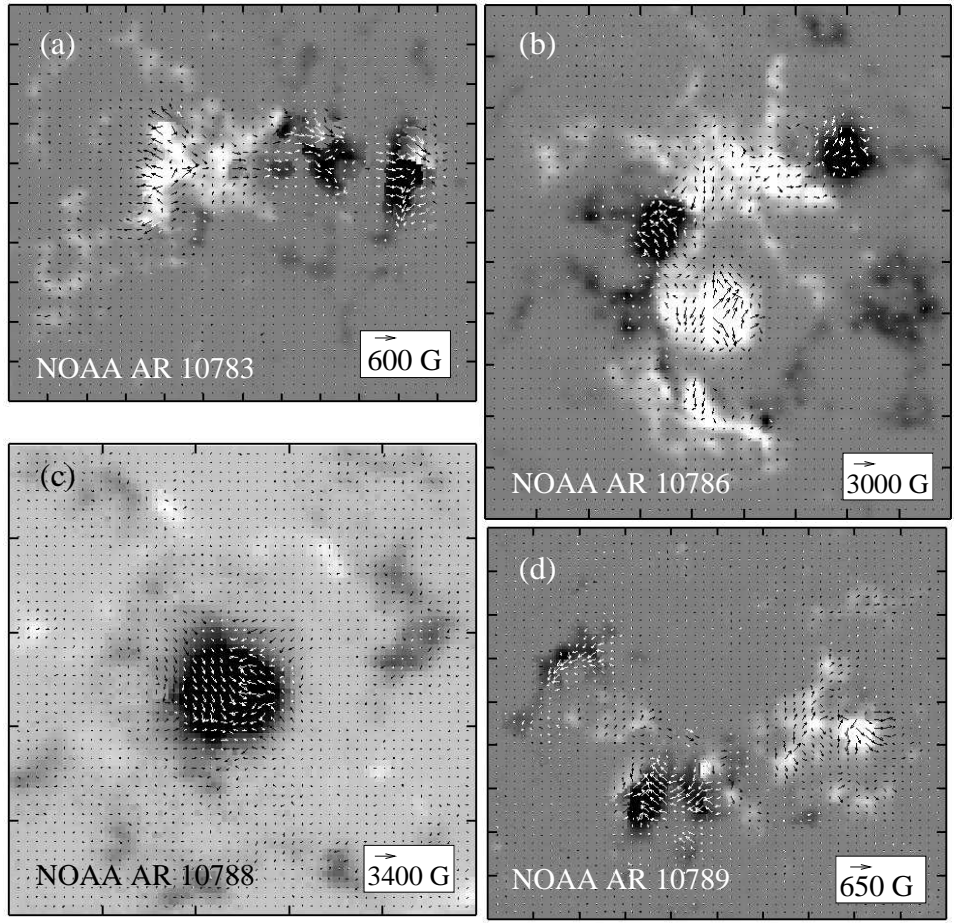}
\caption{\label{f4} 
The NPFC disambiguation solutions for the four active regions
identified in the VSM magnetogram of Figure \ref{f1}.   Shown are the
heliographic magnetic field components on the image plane. The
solutions are given for a part of the area assigned to each region, 
namely, the part containing the majority of the active-region magnetic
flux. Tick mark separation in all images is $20\arcsec$. The
vertical magnetic field in all images saturates at $\pm 1000\;G$.} 
\end{figure}
Figure \ref{f4} provides the NPFC disambiguation solutions for the
active regions identified in the VSM magnetogram of Figure
\ref{f1}. Inspecting Figure \ref{f4}, one might notice 
some inconsistencies in the orientation of the horizontal magnetic
field. These imperfections should be mostly attributed to problems in
the inference of the VSM azimuth angle. Indeed, the VSM 
measurements at the time of the observation (07/07/05) were still
quite preliminary, 
with now-known issues not addressed in the data. The
reason why we chose to show these data, nevertheless, 
is the high degree of
activity in the solar atmosphere at the time of the
observation. Vector magnetograms obtained thereafter showed much less
activity and fewer, simpler, active regions, although the inference of
the azimuth angle was drastically improved. In the near future, when
solar activity will start stepping up toward the next maximum, the VSM
will be fully equipped with sufficient hardware and software to
reliably acquire and process massive amounts of data. 
\section{Conclusions / Current status of the VSM vector magnetogram data}
Although VSM full-disk vector magnetograms at the Fe {\small I}
$630.2\;nm$ photospheric line 
are obtained since August 2003, these data have not
yet been released to the solar and space physics communities. This is
due to work currently underway on the last remaining issues 
that, however, must be addressed prior to releasing any digital data. 
This being said, we are confident that the data will be available for
unrestricted use very soon. 

Complete descriptions of VSM and its data archive can be found at\\
\verb http://solis.nso.edu/solis_data.html. Among other information, 
this web page includes quick-look visualizations of both the latest VSM
vector magnetograms and the regularly updated list of the VSM vector
magnetogram data archive. Featuring a user-friendly interface, one may
view all the magnetic field components for each identified active
region, together with the zenith and azimuth angles of the
disambiguated magnetic field vector. Upon the release of the VSM
data, full Milne-Eddington inversion products will be available nearly
24 hours after the quick-look data acquisition (Henney, Keller, \&
Harvey, 2006). 

As for the active-region identification and azimuth disambiguation
techniques, this article demonstrates that they are general enough to be
applied to both existing and future vector magnetograms, either in
conjunction or independently. Partial-disk magnetograms (e.g., from {\it
Hinode}) might be disambiguated using the NPFC method, while future
SDO/HMI full-disk magnetograms might undergo a process similar to that
described in \S2 for the SOLIS/VSM data.  
\acknowledgements{SOLIS/VSM data used here are produced cooperatively
  by NSF/NSO and NASA/LWS. The National Solar Observatory is operated
  by AURA, Inc. under a cooperative agreement with the National
  Science Foundation. We are grateful to B. W. Lites for 
  providing us with the inverted spectro-polarimeter data from
  Hinode/SOT. Hinode is a Japanese mission developed and
  launched by ISAS/JAXA, with NAOJ as domestic partner and NASA and
  STFC (UK) as international partners. It is operated by these
  agencies in co-operation with ESA and NSC (Norway). The azimuth
  disambiguation and active-region identification studies have 
  received partial support by the NASA LWS TR\&T Grant
  NNG05-GM47G. N.-E. R.'s work is supported by the NSO and NASA Grant
  NNH05AA12I.  
}
\end{document}